\documentclass[preprint]{aastex}

\def\torque{{\cal G}}
\def\momrate{I}
\def\enrate{E}
\def\gapprox{\lower.4ex\hbox{$\;\buildrel >\over{\scriptstyle\sim}\;$}}
\def\lapprox{\lower.4ex\hbox{$\;\buildrel <\over{\scriptstyle\sim}\;$}}
\def\lessthan{\, < \,}
\def\greaterthan{\, > \,}

\def\Max{{\rm Max}}

\slugcomment{accepted by \apj}

\shorttitle{Relativistic Outflows from Accretion Disks}
\shortauthors{Becker, Subramanian, \& Kazanas}

\begin{document}

\title{RELATIVISTIC OUTFLOWS FROM ADVECTION-DOMINATED ACCRETION
DISKS AROUND BLACK HOLES}

\author{Peter A. Becker\altaffilmark{1}}
\affil{Center for Earth Observing and Space Research,
\break George Mason University, Fairfax, VA 22030, USA}

\author{Prasad Subramanian\altaffilmark{2}}
\affil{National Center for Radio Astrophysics, \break
Tata Institute of Fundamental Research, \break
Pune University Campus, Post Bag 3, \break
Ganeshkhind, Pune 411007, India}

\and

\author{Demosthenes Kazanas\altaffilmark{3}} \affil{Laboratory for
High-Energy Astrophysics,
\break NASA Goddard Space Flight Center, Greenbelt, MD
20771, USA}

\altaffiltext{1}{also Department of Physics and Astronomy,
George Mason University, Fairfax, VA 22030, USA; pbecker@gmu.edu}
\altaffiltext{2}{psubrama@ncra.tifr.res.in}
\altaffiltext{3}{kazanas@milkyway.gsfc.nasa.gov}

\begin{abstract} Advection-dominated accretion flows (ADAFs) have
a positive Bernoulli parameter, and are therefore gravitationally
unbound. The Newtonian ADAF model has been generalized recently to
obtain the ADIOS model that includes outflows of energy and angular
momentum, thereby allowing accretion to proceed self-consistently.
However, the utilization of a Newtonian gravitational potential limits
the ability of this model to describe the inner region of the disk,
where any relativistic outflows are likely to originate. In this paper
we modify the ADIOS scenario to incorporate a pseudo-Newtonian
potential, which approximates the effects of general relativity. The
analysis yields a unique, self-similar solution for the structure of the
coupled disk/wind system. Interesting features of the new solution
include the relativistic character of the outflow in the vicinity of the
radius of marginal stability, which represents the inner edge of the
quasi-Keplerian disk in our model. Hence our self-similar solution may
help to explain the origin of relativistic jets in active galaxies. At
large distances the radial dependence of the accretion rate approachs
the unique form $\dot M \propto r^{1/2}$, with an associated density
variation given by $\rho \propto r^{-1}$. This density variation agrees
with that implied by the dependence of the X-ray hard time lags on the
Fourier frequency for a number of accreting galactic black hole
candidates. While intriguing, the predictions made using our
self-similar solution need to be confirmed in the future using a
detailed model that includes a physical description of the energization
mechanism that drives the outflow, which is likely to be powered by the
shear of the underlying accretion disk.

\end{abstract}


\section{INTRODUCTION}

Advection-dominated accretion flows (ADAFs) are attractive models for
explaining the emission from underluminous black hole candidates such as
the Galactic center source. The earliest paper on this subject was by
Ichimaru (1977), and interest in the subject was revived by Narayan \&
Yi (1994, 1995a, 1995b) and by Abramowicz et al. (1995). Comprehensive
reviews of work in this area can be found in Kato et al. (1998) and in
Narayan, Mahadevan, \& Quataert (1998). Briefly, these models describe
the dynamics of gas fed onto a black hole at very low (significantly
sub-Eddington) accretion rates. The gas is optically thin and hot, and
possesses separate ion ($T_i \sim 10^{12}\,$K) and electron ($T_e \sim
10^9\,$K) temperatures because it is too tenuous for Coulomb collisions
to equilibrate the species. The low density also reduces the efficiency
of bremsstrahlung emission, hence limiting the luminosity of the flows.
Although some success has been achieved in understanding underluminous
systems using ADAF models, a number of issues have been raised
concerning their self-consistency. Attention has been focused for
example on the two-temperature assumption, which is one of the basic
features of ADAF models. While some authors have identified conditions
under which this assumption is justified (Blackman 1999; Gruzinov 1998;
Quataert 1998; Quataert \& Gruzinov 1999), the validity of the assumption
has also been challenged based on the possibility that dissipation via
plasma instabilities may preferentially heat the electrons rather than
the protons (Bisnovatyi-Kogan \& Lovelace 1997). Chakrabarti (1999)
has questioned the relevance of the self-similar solutions that
do not incorporate boundary conditions and therefore do not exhibit
shocks.

Narayan, Kato, \& Honma (1997) and Narayan \& Yi (1994, 1995a)
have pointed out that in ADAF models the gas has a positive
Bernoulli parameter and is therefore gravitationally unbound.
The positivity of the Bernoulli parameter reflects the fact
that the viscous torque transports energy as well as angular
momentum outward through the disk, which tends to make the
total local energy per unit mass positive. This could result
in the evaporation and escape of at least some of the gas before
it passes through the event horizon of the black hole. Hence the
standard ADAF models may not be self-consistent, since they assume
that all of the available gas accretes onto the compact object.
This idea has been further supported by the ADAF simulations of
Igumenshchev \& Abramowicz (1999), which indicate that bipolar
outflows may be produced when the Shakura-Sunyaev (1973)
viscosity parameter $\alpha \sim 1$. However, Abramowicz et
al. (2000) have suggested that self-consistent advection-dominated
models with a positive Bernoulli parameter and no outflow can be
constructed if the self-similar approach is abandoned. Ogilvie
(1999) makes a similar point and solves a simplified version
of one-dimensional, quasi-spherical accretion to conclude that
time-dependent accretion can take place without the need for any
outflows. While this is an interesting possibility, it appears that
many underluminous black hole systems such as Sgr~${\rm A}^{*}$
may possess relativistic outflows (Falcke 1999;
Donea et al. 1999 and references therein). We therefore focus
on coupled disk/wind models in this paper.

To address the problem of gravitationally unbound accretion in the
standard ADAF models, Blandford \& Begelman (1999; hereafter BB)
generalized the model to include the possibility of powerful winds that
carry away mass, energy, and angular momentum in proportions that render
the Bernoulli constant of the gas remaining in the disk negative,
thereby allowing it to accrete. The resulting advection-dominated
inflow-outflow solutions (ADIOS) are similar to the ejection-dominated
accretion flows (EDAFs) analyzed by Donea et al. (1999). Beckert (2000)
has also obtained related self-similar solutions for advection-dominated
accretion flows that include outflows. As it stands, the formalism of BB
does not address the critical question of the formation mechanism for
the outflow. This mechanism must be capable of accelerating high-energy
protons, thereby creating a nonthermal tail in the particle
distribution. Protons in the nonthermal tail escape to form the wind
whereas the low-energy protons are tightly bound and therefore accrete.
As discussed by Subramanian, Becker, \& Kazanas (1999; hereafter SBK),
the energy transfer may be accomplished via second-order Fermi
acceleration driven by the shearing magnetic field.

One of the primary reasons for studying black hole accretion is to
understand the connection between accretion and the production of
relativistic jets. The jets probably originate close to the black hole
as relativistic winds that become collimated with increasing distance.
The study of the formation of the wind/jet therefore requires a
reasonably precise treatment of the inner region of the disk, where the
physical conditions are most extreme. Although it provides a useful
means for analyzing steady accretion at large distances from the black
hole, the ADIOS model in its original form is not applicable close to
the event horizon due to the utilization of the standard Newtonian form
for the gravitational potential. In this paper we attempt to incorporate
the effects of general relativity by utilizing the pseudo-Newtonian
approximation for the gravitational potential suggested by Paczy\`nski
\& Wiita (1980). Clearly, this approach is not a substitute for a full
treatment of general relativity, but it does provide us with a
convenient mathematical tool for exploring the structure of the inner
region of the disk. Our strategy is to construct new models for
self-similar, advection-dominated accretion disks and associated winds
that are as faithful as possible to the original ADAF and ADIOS scaling
principles, while incorporating the pseudo-Newtonian form for the
gravitational potential. The solutions obtained describe the structure
of the entire disk outside the radius of marginal stability, which is
the edge of the quasi-Keplerian disk in our model. One clear implication
of our approach is that our model should reduce to the ADIOS form at
large distances from the black hole. We focus here on gasdynamical winds
that exert no torque on the underlying disk. These winds may be powered
by the shear acceleration mechanism discussed by SBK. The remainder of
the paper is organized as follows. In \S~2 we develop the
pseudo-Newtonian analog to the self-similar ADIOS model of BB. The
consequences for the structure of the disk/wind system are analyzed in
\S\S~3 and 4, and the implications for the interpretation of
observations of underluminous black-hole systems are discussed in \S~5.
This includes a consideration of the relevance of the SBK shear
acceleration in the context of the pseudo-Newtonian disk/wind model
developed here.

\section{PSEUDO-NEWTONIAN INFLOW-OUTFLOW MODEL}

Working in units such that $GM=c=1$, we can modify the thin-disk,
quasi-Keplerian ADIOS formalism of BB to approximate the effects of
general relativity close to a Schwarzschild black hole by replacing the
Newtonian potential $\Phi=-1/r$ with the pseudo-Newtonian form
$\Phi=-1/(r-2)$ introduced by Paczy\`nski and Wiita (1980) and further
discussed by Paczy\`nski (1998). This prescription yields the correct
values for the marginally bound and marginally stable orbital radii. The
pseudo-Newtonian potential offers a convenient means of approximating
the effects of general relativity while preserving the degree of
mathematical simplicity contained in the ADIOS solutions. However, as we
demonstrate below, the pseudo-Newtonian potential introduces a length
scale (namely, the horizon at $r=2$), and therefore it is no longer
possible to construct self-similar models based solely on power-law
variations in radius.

\subsection{\it Self-Similar Forms}

Our goal here is to use the same physical principles underlying the
ADIOS scalings to obtain new self-similar forms that are valid in the
inner region of a disk governed by the pseudo-Newtonian potential. The
advantage of the self-similar approach is that the solutions obtained
may be valid over several orders of magnitude in radius. As with any
self-similar model, there is a certain degree of arbitrariness in the
imposed scalings. We therefore base our approach on the same physical
principles employed in the ADIOS model, as applied in the context of the
pseudo-Newtonian potential. We believe that the self-similar solutions
obtained can yield useful insights into possible behaviors that should
be investigated subsequently using models based on rigorous descriptions
of the relevant physics.

BB assumed that the angular velocity of the disk $\Omega$ satisfies
the quasi-Keplerian scaling $\Omega \propto \Omega_K$, where $\Omega_K
\propto r^{-3/2}$ is the Keplerian angular velocity in a Newtonian
potential. In the pseudo-Newtonian case the appropriate expression
for the Keplerian angular velocity is (Paczy\`nski \& Wiita 1980)
$$
\Omega_K(r) = {r^{-1/2} \over r-2} \ , \eqno(2.1)
$$
and therefore the variation of $\Omega$ in a quasi-Keplerian accretion
disk now becomes
$$
\Omega(r) = {v_\phi \over r} = \Omega_0 \, {r^{-1/2} \over r - 2} \,,
\eqno(2.2)
$$
where $\Omega_0$ is a positive constant and $v_\phi$ is the azimuthal
velocity. Hence $\Omega$ scales as $\Omega_K$ throughout the disk. As in
the ADIOS model, we assume that $0 < \Omega_0 < 1$ so that the disk is
sub-Keplerian. Simulations carried out by Narayan et al. (1997) indicate
that the behavior of $\Omega$ is well represented by equation~(2.2) over
several orders of magnitude in radius (outside the radius of marginal
stability) for values of the Shakura-Sunyaev viscosity parameter $\alpha
\sim 0.1 - 0.3$. Based on these results, we shall assume in our work that the
variation of $\Omega$ is given by equation~(2.2). However, this
assumption does not apply in the region below the radius of marginal
stability, where the gas passes through a sonic point before crossing
the horizon. We provide further discussion of this point in \S~5.

Proceeding by analogy with the ADAF/ADIOS models, we shall suppose
that the outwardly-directed energy transport rate $\enrate$ and the
inwardly-directed angular momentum transport rate $\momrate$ in the
disk scale as
$$
\enrate(r) = \epsilon \, {\dot M(r) \over r-2} \ , \ \ \ \ \ \ \ 
\momrate(r) = {\lambda \over \Omega_0} \ \ell \, \dot M(r) \ ,
\eqno(2.3)
$$
respectively, where $\epsilon$ and $\lambda$ are positive constants,
$\ell = r^2 \, \Omega$ is the angular momentum per unit mass, and $\dot
M(r)$ is the positive mass accretion rate. We will also assume that the
internal energy per unit mass (i.e., the ion temperature) scales as the
gravitational potential energy, implying that
$$
a^2(r) \equiv {P \over \rho} = {a_0^2 \over r-2} \ , \eqno(2.4)
$$
where $a$ is the isothermal sound speed, $P$ is the gas pressure, $\rho$
is the mass density, and $a_0$ is a positive constant. All of the
physical quantities denote vertical averages at a given cylindrical
radius $r$. The assumed variations of $\enrate$, $\momrate$, $a$, and
$\Omega$ are the most natural extensions of the scalings used by BB into
the pseudo-Newtonian regime. In \S~3 we shall see that these scalings
impose definite constraints on the mechanism responsible for producing
the energy and mass loss that powers the outflow.

\subsection{\it Disk-Wind Coupling}

Our ultimate goal is to understand the possible role of the shear
acceleration mechanism discussed by SBK in launching ADIOS-type
outflows. Consequently, we focus here on gasdynamical outflows
(winds) that exert no local torque on the disk, and therefore
we stipulate that $\Omega = \Omega_W$, where
$$
\Omega_W \equiv {1 \over r^2} \, {d \momrate \over d \dot M}
= {1 \over r^2} \, {d \momrate / dr \over d \dot M / dr} \eqno(2.5)
$$
denotes the angular velocity of the wind at the point of contact with
the disk. Substituting for $\momrate$ using equation~(2.3) then yields
$$
r^2 \, \Omega \, {d \dot M \over d r}
= {d \over d r} \left(\lambda \, \dot M \, r^2 \,
{\Omega \over \Omega_0} \right) \ , \eqno(2.6)
$$
which immediately implies that the variation of $\dot M$ is given by
$$
\dot M \propto \left(r^2 \, \Omega\right)^{\lambda/(\Omega_0-\lambda)}
\ , \eqno(2.7)
$$
or, using equation~(2.2) for $\Omega(r)$,
$$
\dot M \propto \left(r^{3/2} \over r - 2 \right)^{\lambda/
(\Omega_0-\lambda)} \ . \eqno(2.8)
$$

We can relate the variation of $\dot M$ to the other physical
quantities in the disk by writing
$$
\dot M = 4 \pi \, r \, H \, v \, \rho \ , \eqno(2.9)
$$
where $H$ is the half-thickness of the disk and $v$ is the radial
velocity, defined to be positive for inflow. The value of $H$ can
be characterized by employing the equation of vertical hydrostatic
equilibrium to write
$$
{P \over z} \approx - {\rho \over (r-2)^2} \,
{z \over r} \ , \eqno(2.10)
$$
which is valid provided $z \ll r$. A scale analysis of this expression
yields the usual approximation (e.g., Narayan et al. 1997)
$$
\left(H \over r\right)^2 = a^2 \, {(r-2)^2 \over r}
= a_0^2 \, \left({r-2 \over r}\right) \ , \eqno(2.11)
$$
which reduces to equation~(20) of BB in the limit $r \to \infty$.
Since the disk must be thin for self-consistency, we require that
$a_0 < 1$.

\subsection{\it Radial Momentum Conservation}

Our self-similar approach is based on the scalings for $\enrate$,
$\momrate$, and $a$ given by equations~(2.3) and (2.4), respectively.
We conjecture that the associated radial velocity $v$ scales as
$$
v(r) = {v_0 \over \sqrt{r-2}} \ , \eqno(2.12)
$$
where $v_0$ is a positive constant, implying that the radial velocity
is proportional to the local freefall velocity, $v_{\rm ff}(r) =
\sqrt{2/(r-2)}$. We shall verify this hypothesis by confirming
that the resulting variation of the energy transport rate $\enrate(r)$
satisfies equation~(2.3). Note that in order to avoid inflow faster than
freefall, we require that $v_0^2 < 2$. In a steady state, the radial
equation of motion (eq.~[6] of BB) can be written as
$$
v \, {d v \over d r}
= r \, \Omega^2 - {1 \over (r-2)^2} - {1 \over \rho} \,
{d P \over d r}
\ . \eqno(2.13)
$$
We can obtain a differential equation for the pressure $P$ by
substituting for $\Omega$, $\rho = P/a^2$, and $v$ in equation~(2.13)
using equations~(2.2), (2.4), and (2.12), respectively, which yields
$$
{d \ln P \over d r}
= {v_0^2 + 2 \, \Omega_0^2 - 2 \over 2 a_0^2 (r-2)} \ , \eqno(2.14)
$$
with solution
$$
P(r) = P_0 \, (r-2)^{(v_0^2 + 2 \, \Omega_0^2 - 2)/(2 a_0^2)}
\ , \eqno(2.15)
$$
where $P_0$ is a positive constant. The associated density
variation is given by
$$
\rho(r) = \rho_0 \, (r - 2)^{(v_0^2 + 2 \, \Omega_0^2 +
2 \, a_0^2 - 2)/(2 a_0^2)} \ , \eqno(2.16)
$$
where
$$
\rho_0 \equiv {P_0 \over a_0^2}
\ . \eqno(2.17)
$$

Using equations~(2.11), (2.12), and (2.16) to substitute for $H$,
$v$, and $\rho$, respectively, in equation~(2.9) for the accretion
rate, we obtain
$$
\dot M = \dot M_0 \ r^{3/2} \, (r-2)^{(v_0^2 + 2 \, \Omega_0^2 + 2
\, a_0^2 - 2)/(2 a_0^2)} \ , \eqno(2.18)
$$
where
$$
\dot M_0 \equiv {4 \pi \, a_0 \, v_0 \, \rho_0} \ . \eqno(2.19)
$$
The power-law indices in this expression must match those appearing
in equation~(2.8), and therefore we require that
$$
\lambda = {\Omega_0 \over 2} \ , \ \ \ \ \ \ 
v_0^2 = 2 - 2 \, \Omega_0^2 - 4 \, a_0^2 \ , \eqno(2.20)
$$
which together imply that the self-similar variation of the accretion
rate is given by the unique solution
$$
\dot M(r) = \dot M_0 \, {r^{3/2} \over r - 2} \ . \eqno(2.21)
$$
This solution for $\dot M(r)$ has an interesting behavior, in that it
possesses a minimum at $r=r_{\rm ms} =6$, which is the radius of
marginal stability for a Schwarzschild black hole. The rate at which
matter is fed into the wind therefore approaches zero as $r \to r_{\rm
ms}$. This is physically reasonable since below this radius we expect
that most of the remaining material will be swept directly into the
black hole rather than expelled into the wind. For $r > 6$, $\dot M$
decreases monotonically with decreasing radius in response to the loss
of mass into the wind, and as $r \to \infty$, we find that $\dot M
\propto r^{1/2}$, in agreement with BB provided their parameter $p
\equiv d \ln \dot M /d \ln r = 1/2$. Our sub-Keplerian assumption
$\Omega_0 < 1$ can be combined with the requirement that $v_0^2 > 0$ to
conclude that $a_0^2 < (1 - \Omega_0^2)/2$, confirming that the disk is
thin (see eq.~[2.11]). Incorporating equations~(2.20) into our previous
results, we conclude that the pressure and density vary as
$$
P(r) = {P_0 \over (r-2)^2} \ , \ \ \ \ \ \ \ 
\rho(r) = {\rho_0 \over r-2} \ , \eqno(2.22)
$$
which both increase monotonically with decreasing radius as required.
In \S~5 we discuss the possible relevance of this density distribution
in accounting for the observations of time lags between the soft and
hard X-rays in accreting galactic sources.

\subsection{\it Energy and Angular Momentum Conservation}

Following BB, we express conservation of angular momentum in the
disk by writing the inwardly-directed angular momentum transport
rate as
$$
\momrate(r) = \dot M \, r^2 \, \Omega - \torque \ , \eqno(2.23)
$$
where $\torque(r)$ is the torque exerted by the disk interior to
radius $r$ on the outer portion of the disk. The outwardly-directed
energy transport rate in the disk is likewise given by
$$
\enrate(r) = \torque \, \Omega - \dot M \, \left( {1 \over 2} \,
r^2 \, \Omega^2 + {1 \over 2} \, v^2
+ {5 \over 2} \, a^2 - {1 \over r-2} \right) \ , \eqno(2.24)
$$
which incorporates the pseudo-Newtonian gravitational potential. Note
that equation~(2.24) includes a term describing the radial kinetic
energy of the gas, whereas BB ignored this term in their energy
transport equation. The gas in the disk is assumed to have adiabatic
index $\gamma=5/3$. We can solve for the torque by combining
equations~(2.3) and (2.23) and setting $\lambda = \Omega_0/2$
in compliance with equation~(2.20), which yields
$$
\torque(r) = \momrate(r) =
{1 \over 2} \, \dot M \, r^2 \, \Omega \ . \eqno(2.25)
$$
The physical implications of this torque can be investigated by
combining equation~(2.25) with the expressions for $\dot M$ and $\Omega$
to determine the implied variation of the dynamic viscosity $\nu$. We
discuss this issue further in \S~3. By using equation~(2.25) to
eliminate $\torque$ in equation~(2.24) and substituting for $\Omega$ and
$a$ using equations~(2.2) and (2.4), respectively, we can reexpress the
energy transport rate as
$$
\enrate(r) = \epsilon \, {\dot M(r) \over r-2} \ , \eqno(2.26)
$$
where
$$
\epsilon = 1 - {5 \over 2} \, a_0^2 - {1 \over 2} \, v_0^2
\ . \eqno(2.27)
$$
Equations~(2.20) and (2.27) establish three connections between
the five parameters $\lambda$, $\Omega_0$, $\epsilon$, $a_0$,
and $v_0$. Hence knowledge of any two is sufficient to calculate
the remaining three. We shall therefore treat $\Omega_0$ and
$\epsilon$ as free parameters in our model and compute $\lambda$,
$a_0$, and $v_0$ using the relations
$$
\lambda = {\Omega_0 \over 2} \ , \,\,\,\,\,\,
a_0^2 = 2 \, \Omega_0^2 - 2 \, \epsilon \ , \,\,\,\,\,\,
v_0^2 = 8 \, \epsilon - 10 \, \Omega_0^2 + 2 \ , \eqno(2.28)
$$
obtained using equations~(2.20) and (2.27).

The energy transport rate given by equation~(2.26) does satisfy
equation~(2.3), and therefore we have confirmed the validity of our
solution for $v(r)$ given by equation~(2.12).
The establishment of the self-similar form for
$\enrate$ stems from the fact that in our model, the local torque is
equal to the local angular momentum transport rate (see eq.~[2.25]), and
consequently the outward energy transport due to the torque balances the
inward energy transport due to the accretion of azimuthal kinetic
energy. BB obtain the same result for gasdynamical winds when $p = 1/2$.
The net energy transport is in the outward direction provided $\epsilon
> 0$. As $r \to \infty$, we find that $\enrate(r) \to \epsilon \, \dot M
\, r^{-1}$ in agreement with the ADIOS model of BB. Substituting for
$\dot M$ using equation~(2.21), our results for the energy and angular
momentum transport rates can be rewritten as
$$
\enrate(r) = \epsilon \, \dot M_0 \, {r^{3/2} \over (r-2)^2} \ ,
\ \ \ \ \ \ \ 
\momrate(r) = \lambda \, \dot M_0 \, {r^3 \over (r-2)^2} \ ,
\ \ \ \ \ \ \ 
\torque(r) = \lambda \, \dot M_0 \, {r^3 \over (r-2)^2}
\ . \eqno(2.29)
$$
Note that the torque ${\cal G}$ possesses a minimum at the radius
of marginal stability $r = r_{\rm ms} = 6$, but it does not vanish
there. This behavior is consistent with simulations performed by Hawley
\& Krolik (2000), who find that the viscous stress is finite at
$r = r_{\rm ms}$. Moreover, Narayan et al. (1997) suggest that the
stress actually vanishes at the sonic point (between the horizon
and the radius of marginal stability), where the radial velocity
becomes supersonic and the gas therefore loses pressure support.
It follows that the torque is finite at $r = r_{\rm ms}\,$, as implied
by equation~(2.29). However, it must be pointed out that this issue
is a subject of ongoing debate. While Hawley \& Krolik (2000), Agol
\& Krolik (2000), Gammie (1999) and Krolik (1999) contend that there
is a finite stress at the radius of marginal stability, Armitage
et al. (2000) and Paczynski (2000) argue that this should not be
the case, at least for thin disks. In \S~3 we examine the implications
of our self-similar model for the variation of the viscosity and for
the associated dissipation of kinetic energy in the disk.

\section{\it Viscosity and Dissipation}

Equation~(2.25) for the torque ${\cal G}$ in our self-similar RADIOS
model has definite implications for the variation of the kinematic
viscosity $\nu$. We can calculate the viscosity by writing the shear
stress as (e.g., Frank, King, \& Raine 1985)
$$
{{\cal G} \over 4 \pi r^2 H} = - \rho \, \nu \, r
{d \Omega \over d r} \ , \eqno(3.1)
$$
and eliminating ${\cal G}$ using equation~(2.25), which yields
$$
\nu = - {v \, \Omega \over 2} \left({d \Omega \over d r}
\right)^{-1} \ . \eqno(3.2)
$$
Substituting for $\Omega$ and $v$ using equations~(2.2) and (2.12),
respectively, we obtain
$$
\nu = {v_0 \, r \, (r-2)^{1/2} \over 3 r - 2} \ . \eqno(3.3)
$$
We can evaluate the variation of $\nu$ most straightforwardly by
calculating an equivalent value for the associated Shakura-Sunyaev
$\alpha$-parameter, defined by writing the shear stress as
$$
\alpha \, P \equiv - \rho \, \nu \, r \, {d\Omega \over dr}
\ . \eqno(3.4)
$$
Setting $P = \rho a^2$ and combining the result with equations~(2.2),
(2.4), (2.28), and (3.3) yields
$$
\alpha = \alpha_0 \, \left({r \over r-2}
\right)^{1/2} \ , \eqno(3.5)
$$
where
$$
\alpha_0 \equiv {\Omega_0 \, (8 \, \epsilon - 10 \,
\Omega_0^2 + 2)^{1/2}
\over 4 \, (\Omega_0^2 - \epsilon)}
\ . \eqno(3.6)
$$
We plot $\alpha_0$ as a function of $\epsilon$ and $\Omega_0$ in
Figure~1. Equation~(3.5) indicates that $\alpha$ has a very weak radial
dependence, increasing by a factor of $1.2$ as $r$ decreases from
infinity to $r=r_{\rm ms}=6$. Hence the $\alpha$-parameter associated
with our self-similar RADIOS model is practically constant within the
hot inner region of the disk, which is consistent with the conventional
assumption that the shear stress is approximately proportional to the
pressure (e.g., Narayan et al. 1997). Based upon the observed behavior
of $\alpha$, we conclude that our model has an implied viscosity
variation that is consistent with fundamental notions regarding
the transport of angular momentum in the shear flow. This tends to
further support our fundamental assumptions regarding the scalings of
$\enrate$, $\momrate$, and $a$.

\subsection{\it Implications for Dissipation and Cooling}

Next we consider the implications of our self-similar model for
the dissipation on energy in the shear flow, and for the rate
of cooling from the disk due to the loss of energetic particles
into the wind. The equation governing the internal energy of
the matter in the disk is
$$
{DU \over Dt} = \gamma \, {U \over \rho} \, {D\rho \over Dt}
+ \dot U_{\rm viscous} - \dot U_{\rm wind} \ , \eqno(3.7)
$$
where $D/Dt$ represents the comoving (Lagrangian) time derivative
following the gas, $\gamma$ is the adiabatic index, $U=P/(\gamma-1)$ is
the internal energy density, and $\dot U_{\rm viscous}$ and $\dot U_{\rm
wind}$ denote the (positive) rates of viscous heating and cooling into
the wind, respectively. Using the standard expression for the viscous
heating rate (Frank, King, \& Raine 1985),
$$
\dot U_{\rm viscous} = \rho \, \nu \, r^2 \, \left(
{d\Omega \over dr}\right)^2 \ , \eqno(3.8)
$$
we obtain in a steady state
$$
{v \, \rho \over 1-\gamma} \, {da^2 \over dr}
+v \, a^2 \, {d\rho \over dr} = \rho \, \nu \, r^2 \, \left(
{d\Omega \over dr}\right)^2 - \dot U_{\rm wind} \ . \eqno(3.9)
$$
This is equivalent to equation~(2.12) from Narayan et al. (1997), except
that we have chosen $v$ to be positive for inflow. In the standard ADAF
theory, a factor $f$ is introduced to represent the fraction of the
energy dissipated via viscosity that goes into heating the gas, with the
remainder escaping as radiative losses. We can define an analogous
heating efficiency factor $f$ in the context of our disk/wind model
by writing
$$
f \, \rho \, \nu \, r^2 \, \left(
{d\Omega \over dr}\right)^2 \equiv \rho \, \nu \, r^2 \, \left(
{d\Omega \over dr}\right)^2 - \dot U_{\rm wind}
\ . \eqno(3.10)
$$
Hence $1-f$ gives the fraction of the energy dissipated via viscosity
that is expelled into the wind via the internal energy of the escaping
particles. Narayan et al. and other authors focusing on the standard
ADAF scenario usually assume that $f=1$ throughout the disk. However, in
our RADIOS model we can calculate $f$ self-consistently by solving for
$\dot U_{\rm wind}$ and then utilizing equation~(3.10). Setting
$\gamma=5/3$ and substituting for $v$, $a^2$, $\rho$, $\nu$, and
$\Omega$ using our self-similar model, we find that equation~(3.9)
yields for the implied wind cooling rate
$$
\dot U_{\rm wind} = {\rho_0 \, v_0 \over 4}
\left[ {\Omega_0^2 \, (6-r) + 4 \, \epsilon \, (r-2) \over
(r-2)^{9/2}}\right]
\ . \eqno(3.11)
$$
The corresponding result for the heating efficiency factor obtained
using equation~(3.10) is
$$
f(r) = 4 \left({\Omega_0^2 - \epsilon \over \Omega_0^2}\right)
\left({r - 2 \over 3 \, r - 2}\right) \ . \eqno(3.12)
$$
Since the energy dissipated from the shear flow via viscosity is
divided between heating of the thermal particles and acceleration
of the nonthermal (wind) particles, it follows that $f$ must lie
in the range $0 < f < 1$ if we are to obtain a physically reasonable
solution. In order to satisfy this condition as $r \to \infty$, we
must stipulate that
$$
\epsilon \greaterthan {\Omega_0^2 \over 4} \ .
\eqno(3.13)
$$
Note that the heating efficiency $f$ displays a weak radial dependence,
increasing by a factor of $4/3$ as $r$ varies from $r=6$ to infinity.
This indicates that a nearly fixed fraction
of the energy supplied via the dissipation of the shear goes into
heating of the disk, with the remainder powering the outflow. This
seems to be a reasonable result, and one cannot say more in the
absence of a detailed model for the acceleration mechanism
that energizes the wind particles.
We plan to address this issue by developing
an advection-dominated disk/wind model that incorporates the
Fermi-shear acceleration process discussed by SBK in future work.

\subsection{\it Parameter Constraints}

We have assumed that the disk is thin and sub-Keplerian, and that
the inflow velocity is less than the local freefall velocity. These
assumptions can be combined to constrain the range of values for the
various parameters appearing in the model. In order for the disk to
be thin, we require that $a_0^2 < 1$. Equation~(2.28) for $a_0^2$
then implies that for a given value of $\Omega_0$, the parameter
$\epsilon$ must satisfy
$$
\Max\left(0 \ , \ \Omega_0^2 - {1 \over 2}\right)
\lessthan \epsilon \lessthan \Omega_0^2
\ , \eqno(3.14)
$$
where the upper bound follows from the requirement that $a_0^2 > 0$.
By combining equation~(3.14) with our sub-Keplerian restriction
$\Omega_0 < 1\,$, we find that $\epsilon$ cannot exceed unity
in general. We must also stipulate that $0 < v_0^2 < 2$ in order
to keep $v^2$ positive while avoiding inflow faster than the
free-fall velocity. Equation~(2.28) for $v_0^2$ then implies
that $\epsilon$ must fall within the range
$$
\Max\left(0 \ , \ {5 \, \Omega_0^2 - 1 \over 4}\right)
\lessthan \epsilon \lessthan
{5 \, \Omega_0^2 \over 4} \ . \eqno(3.15)
$$
We plot the constraints represented by equations~(3.14) and (3.15)
in the $(\Omega_0 \, , \epsilon)$ parameter space in Figure~2. The
intersection of the constraints is expressed by the condition
$$
\Max\left(0 \ , \ {5 \, \Omega_0^2 - 1 \over 4}\right)
\lessthan \epsilon \lessthan
\Omega_0^2 \ . \eqno(3.16)
$$
We can combine this condition with equation~(3.13) to conclude that
in order for our model to be {\it fully} self-consistent, $\epsilon$
and $\Omega_0$ must satisfy
$$
\Max\left({\Omega_0^2 \over 4} \ , \ {5 \, \Omega_0^2 - 1 \over 4}\right)
\lessthan \epsilon
\lessthan
\Omega_0^2
\ , \ \ \ \ \ \ 
0 \lessthan \Omega_0^2 \lessthan 1
\ . \eqno(3.17)
$$
The allowed region of the $(\Omega_0 \, , \epsilon)$ parameter
space is the curved, shaded region depicted in Figure~2. For
convenience, we have
also included the shaded region of self-consistency in Figure~1
for $\alpha_0$. Note that we can obtain values for
$\alpha_0$ in the range $0 < \alpha_0 < 1$ within the region
of the parameter space satisfying the conditions
given by equations~(3.17). In \S~4 we present a quantitative
discussion of the implications of the pseudo-Newtonian potential
for the coupled disk/wind structure.

\section{EFFECTS OF THE RELATIVISTIC POTENTIAL}

In this section we compare the results obtained using our self-similar,
pseudo-Newtonian, relativistic advection-dominated inflow-outflow
solution (RADIOS) with the ADIOS results derived by BB using the
standard Newtonian potential. Before proceeding, it is important to
point out that the ADIOS model includes the possibility of a mismatch
between $\Omega$ and $\Omega_W$, i.e., there may be a torque between the
disk and the wind. Since the winds treated here are gasdynamical, we
have assumed that no such torque exists, and consequently we have set
$\Omega=\Omega_W$. This corresponds to case (iv) of BB, with
$\lambda = \lambda_{\rm crit}$, where
$$
\lambda_{\rm crit} =
2p \left[ {(10\epsilon + 4p
-4\epsilon p) \over (2p+1) \, (4p^2 + 8p +15)} \right]^{1/2}
\ . \eqno(4.1)
$$
Furthermore, as we have already pointed out, our utilization of the
pseudo-Newtonian potential imposes a length scale (i.e., the horizon at
$r=2$) that implies a definite behavior for $\dot M$ as given by
equation~(2.21). This in turn requires that we set $p = 1/2$ in the
ADIOS model in order to match the behavior of our $\dot M$ solution as
$r \to \infty$. Incorporating this value for $p$ into equation~(4.1)
then yields
$$
\lambda = {1 \over 2} \, \left(1 + 4 \epsilon \over 5
\right)^{1/2} \ , \eqno(4.2)
$$
which can be combined with equations~(19) and (20) of BB to obtain
$$
\Omega_0 = \left(1 + 4 \epsilon \over 5
\right)^{1/2} \ , \ \ \ \ \ \ \ 
a_0 = \left(2 - 2 \epsilon \over 5 \right)^{1/2}
\ . \eqno(4.3)
$$
These results are consistent with our equations~(2.28) in the special
case $v_0 = 0$, which is due to the fact that BB neglected the radial
kinetic energy term in their energy transport equation. We shall use the
subscript ``$A$'' to denote quantities related to the ADIOS model.
Unsubscripted quantities will continue to refer to our RADIOS model. The
self-similar ADIOS expressions for the angular velocity and the
isothermal spound speed are then given by
$$
\Omega_A(r) \equiv \Omega_0 \, r^{-3/2} \ , \ \ \ \ \ \ \ 
a_A(r) \equiv a_0 \, r^{-1/2} \ , \eqno(4.4)
$$
respectively. The corresponding ADIOS solutions for the energy,
angular momentum, and mass transport rates reduce when $p = 1/2$ to
$$
\enrate_A(r) \equiv  \epsilon \, \dot M_0 \, r^{-1/2}\ , \ \ \ \ \ \ \ 
\momrate_A(r) \equiv \lambda \, \dot M_0 \, r \ , \ \ \ \ \ \ \ 
M_A(r) \equiv \dot M_0 \, r^{1/2} \ , \eqno(4.5)
$$
respectively. These solutions obviously agree with our equations~(2.21)
and (2.29) in the limit $r \to \infty$ as required. In Figures~3 and 4
we compare our results for the energy and mass transport rates
$\enrate(r)$ and $\dot M(r)$ with the corresponding ADIOS quantities
$\enrate_A(r)$ and $\dot M_A(r)$ for radii in the range $r_{\rm ms} < r
< 10$, where $r_{\rm ms}=6$. Note that for the same value of $\epsilon$,
energy is transported outward through the disk at a substantially
greater rate in the relativistic (pseudo-Newtonian) model due to the
steepness of the potential in the inner region of the disk.

\subsection{\it Disk/Wind Structure}

The modifications to the ADIOS disk structure due to the incorporation
of the pseudo-Newtonian gravitational potential in RADIOS can be
summarized by writing
$$
{\dot M \over \dot M_A} =
{\Omega \over \Omega_A} =
\left( a \over a_A \right)^2 =
\left( \momrate \over \momrate_A \right)^{1/2} =
\left( \enrate \over \enrate_A \right)^{1/2} =
{r \over r-2} \ ,
\eqno(4.6)
$$
which together suggest that no dramatic changes in the disk properties
occur until one approaches the horizon at $r=2$. In fact, the importance
of the subtle variations in structure between the relativistic and
nonrelativistic models really becomes clear only when one examines the
properties of the associated outflows. Due to the symmetry across the
disk midplane, identical outflows emanate from the upper and lower
surfaces of the disk. For brevity, we shall refer to these two outflows
collectively as the ``wind.'' The total amounts of mass and energy
emitted into the wind per unit time between radii $r_{\rm ms}$ and $r$
are given by
$$
\dot m(r) = \dot M(r) - \dot M(r_{\rm ms}) \ , \ \ \ \ \ \ \ 
L(r) = \enrate(r_{\rm ms}) - \enrate(r) \ , \eqno(4.7)
$$
for the pseudo-Newtonian model. The corresponding expressions in
the Newtonian case are
$$
\dot m_A(r) = \dot M_A(r) - \dot M_A(r_{\rm ms}) \ , \ \ \ \ \ \ \ 
L_A(r) = \enrate_A(r_{\rm ms}) - \enrate_A(r) \ . \eqno(4.8)
$$
In the absence of any outflows ($\epsilon=0$), the energy and momentum
transport rates are conserved and therefore $L(r)=L_A(r)=0$, as in the
ADAF models of Narayan \& Yi (1994). At large radii, we find that the
total power emitted into the wind approaches
$$
L(\infty) \equiv \enrate(r_{\rm ms})
= \epsilon \, \dot M_0 \, {6^{3/2} \over 16} \ , \eqno(4.9)
$$
and
$$
L_A(\infty) \equiv \enrate_A(r_{\rm ms})
= \epsilon \, \dot M_0 \, 6^{-1/2}\ , \eqno(4.10)
$$
for the pseudo-Newtonian and Newtonian models, respectively. If we
interpret the radius of marginal stability $r_{\rm ms}=6$ as the inner
edge of the disk, then the values $L(\infty)$ and $L_A(\infty)$
represent the total power emerging from the disk into the wind from all
radii $r > 6$. Equations~(4.9) and (4.10) indicate that
$L(\infty)/L_A(\infty) = 2.25$, so that the relativistic potential
roughly doubles the total power in the outflow for fixed $\epsilon$ and
$\dot M_0$. In an actual accretion disk, we expect that the wind will
terminate at some outer radius $r_{\rm out} > r$ at which the hot inner
region transitions into the optically thick, cool outer disk. We are not
particularly concerned with the value of $r_{\rm out}$ here, although we
note that most hot disk models give values in the range $r_{\rm out}
\sim 10^{2-4}$ (Narayan, McClintock, \& Yi 1996; Narayan, Barret, \&
McClintock 1997).

It is also interesting to compute the amount of mass and energy
emitted into the wind per unit radius per unit time. Utilizing our
pseudo-Newtonian RADIOS expressions yields
$$
{d L \over d r}
= \epsilon \, \dot M_0 \, {r^{1/2} \, (r+6)
\over 2 \, (r-2)^3} \ , \ \ \ \ \ \ \ 
{d \dot m \over d r}
= \dot M_0 \, {r^{1/2} \, (r-6)
\over 2 \, (r-2)^2} \ . \eqno(4.11)
$$
The corresponding expressions obtained using the standard ADIOS
model of BB are
$$
{d L_A \over d r}
= \epsilon \, \dot M_0 \, {r^{-3/2} \over 2} \ , \ \ \ \ \ \ \ 
{d \dot m_A \over d r}
= \dot M_0 \, {r^{-1/2} \over 2} \ . \eqno(4.12)
$$
These results are compared graphically in Figures~5 and 6 for $r$ in the
range $6 < r < 10$. For a given value of $\epsilon$, the power emitted
into the wind is significantly greater in the pseudo-Newtonian case than
in the standard Newtonian model due to the steeper potential gradient.
Note that in the pseudo-Newtonian case, $d \dot m / d r=0$ at the radius
of marginal stability $r_{\rm ms}=6$. As explained earlier, we therefore
interpret $r_{\rm ms}$ as the inner edge of the disk, and assume that
essentially all of the matter remaining at that radius falls directly
into the black hole.

\subsection{\it Bernoulli Parameters}

Another useful way of contrasting the properties of the Newtonian and
pseudo-Newtonian inflow-outflow models is to calculate the specific energy
(i.e., the Bernoulli parameter) in the disk and the wind as functions
of radius. In the pseudo-Newtonian case, the disk Bernoulli parameter
is given by
$$
B(r) = {1 \over 2} \, v^2 + {1 \over 2} \, r^2 \, \Omega^2
+ {5 \over 2} \, a^2 - {1 \over r-2}
= {20 \Omega_0^2 - 12 \epsilon - 4 + (2 + 6 \epsilon - 9 \Omega_0^2) \, r
\over 2 (r-2)^2}
\ , \eqno(4.13)
$$
where we have utilized equations~(2.28) to arrive at the final
result. The corresponding expression in the Newtonian case is
$$
B_A(r) = {1 \over 2} \, r^2 \, \Omega_A^2
+ {5 \over 2} \, a_A^2 - {1 \over r}
= {1 - 6 \epsilon \over
10 \, r} \ , \eqno(4.14)
$$
obtained using equations~(4.3) and (4.4). In equations~(4.13) and
(4.14), $r$ is interpreted as the cylindrical ``starting radius'' at
the base of the wind in the disk. After leaving the disk, the wind is
expected to expand to larger cylindrical radii as the outflow proceeds.
Note that equation~(4.13) for $B$ includes the contribution due to the
radial kinetic energy, whereas the expression for $B_A$ employed by BB
(eq.~[4.14]) assumes that $v_0=0$.

In general, the pseudo-Newtonian disk Bernoulli parameter $B$
may be either positive or negative depending on the values of
$\epsilon$, $\Omega_0$, and $r$. For given values of $\Omega_0$
and $r$, we find that $B < 0$ if and only if $\epsilon < \epsilon_*$,
where
$$
\epsilon_*
\equiv {4 - 2 r + (9 r - 20) \, \Omega_0^2
\over 6 r - 12}
\ . \eqno(4.15)
$$
The roots of $B$ occur at radius $r = r_*$, where
$$
r_* \equiv {12 \, \epsilon - 20 \, \Omega_0^2 + 4 \over
2 + 6 \, \epsilon - 9 \, \Omega_0^2} \ , \eqno(4.16)
$$
with $B < 0$ for $r > r_*$ and $B > 0$ for $r < r_*$. The situation
is complicated by the fact that $\epsilon$ and $\Omega_0$ must also
satisfy the self-consistency conditions given by equations~(3.17).
However, we can make a few general statements; (i) If $\Omega_0^2 < 1/3$,
then $B > 0$ for all radii $r > 6$; (ii) If $1/3 < \Omega_0^2 < 4/5$,
then there is an allowed range of $\epsilon$ values for which $B$
changes sign at $r = r_*$, where $r_* > 6$; (iii) If $\Omega_0^2 > 4/5$,
then $B < 0$ for all radii $r > 6$. Hence, in the pseudo-Newtonian case,
the inner region of the disk may be unbound even when the outer disk
is bound.

In Figure~7 we plot $B$ and $B_A$ as functions of radius for various
values of $\epsilon$, with $v_0 = 0$ in order to facilitate a direct
comparison between the RADIOS and ADIOS model results. Recall that
setting $v_0 = 0$ implies that $\Omega_0 = \sqrt{(4 \, \epsilon + 1)/5}$
according to the last of equations~(2.28). In the Newtonian case, we
find that $B_A > 0$ for all values of $r$ if $\epsilon < 1/6$, and $B_A
< 0$ for all radii if $\epsilon > 1/6$. In the pseudo-Newtonian case
with $v_0 = 0$, we also find that $B > 0$ for all values of $r$ if
$\epsilon < 1/6$, but if $\epsilon > 1/6$, then $B$ changes sign at $r_*
= 20 \, \epsilon / (6 \, \epsilon -1)$. Note that $B$ exceeds $B_A$ at
all radii, which is due to the effect of the pseudo-Newtonian potential.

A positive disk Bernoulli parameter violates self-consistency in
standard ADAF models because these do not include outflows. By contrast,
in the ADIOS and RADIOS models considered here, outflows are included
and therefore we can obtain self-consistent solutions even when the
Bernoulli parameter is positive, although the most plausible scenario
is one in which it is driven to a negative value as a result of
the outflow. Specifically, the most energetic outflows are obtained
when $\epsilon = 1$, which yields a negative (bound) value for the disk
Bernoulli parameter for all values of $r$. Conversely,
the most positive values for the
Bernoulli parameter are obtained when $\epsilon = 0$, which corresponds
to the no-wind solution since in this case the energy transport rate in
the disk vanishes. The source of the change in
behavior of the RADIOS results plotted in Figure~7 relative to those
obtained using the ADIOS model can be traced to the quasi-Keplerian form
for $\Omega$ operative in the pseudo-Newtonian case, expressed by
equation~(2.2). As a consequence of equation~(2.2), the azimuthal
kinetic energy term $(1/2)r^2 \, \Omega^2$ in equation~(4.13) for $B$
varies as $r / (r - 2)^2$, whereas the terms describing the internal and
gravitational potential energies each vary as $1/(r-2)$. At sufficiently
small radii, the azimuthal kinetic energy (which is positive) therefore
overwhelms the remaining terms, which have the weaker $1/(r-2)$ radial
dependence. The vigorous azimuthal motion is due to the stronger
gradient in the pseudo-Newtonian potential relative to the Newtonian
potential. In the ADIOS model all three terms in equation~(4.14) for
$B_A$ have the same scaling, and therefore $B_A$ has the same sign at
all radii.

The Bernoulli parameter for the wind particles is given in
the RADIOS model by
$$
b(r) \equiv {d L \over d \dot m}
= - {d \enrate / d r \over d \dot M / d r}
= {\epsilon \, (r+6) \over (r-6) \, (r-2)}
\ . \eqno(4.17)
$$
The equivalent expression obtained using the ADIOS model results is
$$
b_A(r) \equiv {d L_A \over d \dot m_A}
= - {d \enrate_A / d r \over d \dot M_A / d r}
= {\epsilon \over r}
\ . \eqno(4.18)
$$
Note that $b$ and $b_A$ are always positive, and that they each exceed
the associated disk Bernoulli parameters $B$ and $B_A$, respectively.
This implies that the energy per unit mass in the wind is higher than
that in the disk, which is consistent with our disk/wind scenario.
As the wind flows away from the black hole, we expect the gas to expand
adiabatically, passing through a critical point at which the flow
velocity equals the sound speed as discussed by SBK. In this situation,
the specific energy of the wind particles is conserved and resides
ultimately in the kinetic energy of the outflow. It must be emphasized
that the wind Bernoulli parameter describes only the kinetic, potential,
and internal energy per unit mass, since it is based on the energy
transport equation in the disk, which is classical in form (see
eq.~[2.24]). Hence the rest mass energy of the wind particles is not
included in either $b$ or $b_A$. It follows that if the outflowing gas
expands adiabatically, then the total energy per unit mass in the cold,
terminal wind is given in the pseudo-Newtonian and Newtonian cases by
(e.g., SBK)
$$
\Gamma \approx b + 1 \ , \ \ \ \ \ \ 
\Gamma_A \approx b_A + 1 \ , \eqno(4.19)
$$
respectively, where $\Gamma$ and $\Gamma_A$ are the corresponding
asymptotic Lorentz factors of the outflow in the two models.

Without considering the structure of the outflow in any detail,
we can use equation~(4.19) to determine the asymptotic nature
of the wind originating at radius $r$. In the pseudo-Newtonian
RADIOS case we can combine equations~(4.17) and (4.19) to obtain
$$
\Gamma(r) \approx b(r) + 1
= {\epsilon \, (r+6) \over (r-6) \, (r-2)} + 1
\ . \eqno(4.20)
$$
In the Newtonian ADIOS case the corresponding
result obtained using equations~(4.18) and (4.19) is
$$
\Gamma_A(r) \approx b_A(r) + 1
= {\epsilon \over r} + 1
\ . \eqno(4.21)
$$
According to equations~(3.17), the maximum value of $\epsilon$
satisfying our self-consistency conditions is $\epsilon = 1$,
corresponding to the limiting case $\Omega_0 = 1$. We find that in the
pseudo-Newtonian RADIOS model with $\epsilon = 1$, the asymptotic
Lorentz factor $\Gamma \gapprox 2 $ for $r \lapprox 8.3$, and $\Gamma
\gapprox 3 $ for $r \lapprox 7.3$. This indicates that the inner region
of the disk can produce a relativistic wind/outflow that may
subsequently transition into a relativistic jet at large distances from
the black hole, provided the gas expands adiabatically. Conversely, in
the corresponding ADIOS model with $\epsilon = 1$, the largest possible
value for $\Gamma_A$ is $1.17$ (obtained for starting radius $r = r_{\rm
ms}=6$), and therefore the standard ADIOS outflow is clearly
nonrelativistic. These results are depicted in Figure~8.

\section{DISCUSSION AND CONCLUSION}

Our relativistic, advection-dominated inflow-outflow solution (RADIOS)
incorporates a pseudo-Newtonian gravitational potential, and therefore
it provides an approximate representation of the effects of general
relativity on the properties of the accretion disk and the associated
outflow, which is assumed here to be a gasdynamical wind. Consideration
of mass, momentum, and energy conservation yields a new, self-similar
solution for the disk/wind structure that is rather different from that
obtained by BB. In particular, our results show that a powerful
relativistic outflow originates from the inner region of the disk, close
to the radius of marginal stability. The appearance of this outflow is a
consequence of the steep gradient in the pseudo-Newtonian potential.
Consequently, the RADIOS model may provide a better description of the
inner region of the disk than the standard ADIOS model, which is based
on Newtonian gravitation. Although it utilizes a pseudo-Newtonian
potential, our model nonetheless incorporates the same fundamental
physical scaling principles as the ADIOS model. Namely, we assume that
the angular velocity $\Omega$ scales as the Keplerian value $\Omega_K$
(see eqs.~[2.1] and [2.2]), and we assume that the energy transport rate
per unit mass $\enrate / \dot M$ and the specific internal energy $a^2$
each scale as the gravitational potential energy $\Phi = -1/(r-2)$.
We also assume that the angular momentum transport rate per unit mass
$I / \dot M$ scales as the specific angular momentum $r^2 \, \Omega$.

Evidence supporting our assumption regarding the variation of $\Omega$
is provided by the simulations of Narayan et al. (1997), who recover
this behavior over several orders of magnitude in radius (outside the
radius of marginal stability) for $\alpha \sim 0.1 - 0.3$. Although
Narayan et al. (1997) assume that there is no outflow from the disk,
we can still utilize our equation~(2.2) based on their work because we
assume here that there is no relative torque between the disk and the
wind. Hence we expect little change in the angular momentum distribution
as a result of the outflow. This also supports our assumption regarding
the variation of the angular momentum transport rate $\momrate$ since in
a self-similar model such as ours, the scaling of $\momrate$ follows
directly from the variation of $\Omega$ by virtue of equation~(2.23).

The pseudo-Newtonian potential introduces a length scale into
the problem due to the existence of the horizon at $r=2$. No
such length scale appears in the Newtonian ADIOS model of BB.
The imposition of a length scale in our model results in a definite
behavior for the variation of the accretion rate given by
equation~(2.21), which approaches $\dot M \propto r^{1/2}$ as
$r \to \infty$. When coupled with our assumption of a torque-free
(gasdynamical) wind, the number of free parameters in our model
is reduced to two, which we take to be $\Omega_0$ and $\epsilon$.
The remaining constants $\lambda$, $a_0$, and $v_0$ can be computed
in terms of $\Omega_0$ and $\epsilon$ using equations~(2.28). The
self-similar solutions we obtain asymptotically approach the
corresponding, Newtonian ADIOS results for a gasdynamical wind
as $r \to \infty$ if we set $p \equiv d \ln \dot M /d \ln r = 1/2$
and $\lambda = \lambda_{\rm crit}$ in the ADIOS model (see eq.~[4.1]).
Hence the gravitational physics near the horizon forces the flow
structure to satisfy $\dot M \propto r^{1/2}$ at large radii. In
the Newtonian ADIOS model satisfying this constraint, the outflow
is always nonrelativistic. In our model, the outflow can be highly
relativistic, implying a possible connection with large-scale jets.

The asymptotic Lorentz factor in the wind, $\Gamma$, formally diverges
as $r \to 6$ because $d \dot m /d r$ vanishes at the radius of marginal
stability $r_{\rm ms}=6$. This is due to the fact that for $r < 6$, the
disk is unstable and all of the remaining material is swept directly
into the black hole (see, however, our discussion of the implications of
transonic flow below). Conversely, the differential power in the
outflow, $d L/dr$, remains finite even at $r = 6$ because energy
continues to be transported in the outward radial direction by the
self-similar torque, which has a radial dependence given by
equation~(2.25). Even with our self-consistency conditions (eqs.~[3.17])
taken into consideration, equation~(4.20) indicates that the asymptotic
Lorentz factor for the outflow $\Gamma$ can still greatly exceed unity
for the portion of the wind emitted close to the radius of marginal
stability. By contrast, when the same parameters are used in the ADIOS
model of BB, the outflow is subrelativistic from all radii in the disk
(cf. eq.~[4.21]).

The radial Mach number is conserved in our self-similar model, and
therefore the flow remains subsonic in the entire computational domain
($r > r_{\rm ms}$). Of course, in reality the gas must certainly achieve
a supersonic velocity before it crosses the event horizon. The
transition to supersonic flow requires that the material pass smoothly
through a sonic point, which is a critical point for the flow. The
existence of the sonic point therefore imposes critical conditions that
the overall dynamical structure must satisfy. This does not appear
to be a major problem for our model because the sonic point is expected
to lie between $r = r_{\rm ms}$ and the horizon $r = 2$, which is
outside the computational domain of our disk model. However, the
issue can be addressed quantitatively by constructing a {\it global} model in which the
subsonic region ($r > r_{\rm ms}$) is described by our self-similar
RADIOS solution and the transonic region ($2 < r < r_{\rm ms}$) is
described by the solution of Narayan et al. (1997; see also Chen et al.
1997). A more sophisticated alternative would be to modify the Narayan
et al. calculation to include mass loss such as that envisioned here. We
expect that in the resulting model, matter will actually be expelled
into the wind all the way down to the sonic radius (located between the
radius of marginal stability and the horizon), where the pressure
support for the wind essentially vanishes. Hence, there should still be
some minimal amount of outflow at $r=r_{\rm ms}$, and consequently
$\Gamma$ should remain finite at the radius of marginal stability, in
contrast to our results. Nonetheless, we expect that the dynamical structure
of such a generalized model will be similar to that of the RADIOS model
because the $\Omega(r)$ distribution that we utilize is consistent with
that obtained by Narayan et al. We therefore believe that our model
provides an intriguing glimpse into the possible behavior to be expected
when fully relativistic calculations that include particle acceleration
and mass loss become available.

\subsection{\it Shear Acceleration and the Origin of the Outflows}

In the treaments of advection dominated inflow-outflow solutions that
have appeared to date in the literature (BB; Donea et al. 1999), the
existence of outflows is generally assumed without any elaboration on the
physical mechanisms that could give rise to them. That is also the case
in the work presented here, since the outflow power is calculated by
differentiating the disk energy transport rate $\enrate$, without making
any reference to a microphysical model. Although several mechanisms for
producing outflows have been proposed (see references in SBK), few make
a connection between the outflow and the underlying accretion disk. Das
\& Chakrabarti (1999) and Das (1999) describe pressure-driven winds from
centrifugally-supported boundary layers and shocks in the inner regions
of disks. Xu \& Chen (1997) proposed an advection-dominated flow where
the central object effectively acts as a scatterer which redirects the
inward flow at low latitudes into an outflow at high latitudes. The
above-mentioned models compute the mass outflow rates self-consistently,
but the associated winds are subrelativistic and therefore the possible
connection with large-scale relativistic jets is unclear.

The formation of a relativistic outflow requires the operation of a
mechanism in the disk that preferentially accelerates high energy
particles. The resulting particle distribution comprises a thermal
component that is bound to the disk and a nonthermal component that is
unbound, representing the particles that are destined to escape into the
wind. SBK have described a second-order Fermi acceleration mechanism
powered by the shearing magnetic field that is capable of producing a
relativistic outflow. Some related conceptual development has also been
carried out by Blackman (1999) and by Gruzinov \& Quataert (1999). The
physical scenario in which the acceleration of the wind protons takes
place is the same as that in which hybrid viscosity (Subramanian,
Becker, \& Kafatos 1996) is operative. The azimuthal kinetic energy in
the shear flow is dissipated partly by viscous heating of the thermal
protons, and partly by second-order Fermi acceleration of the nonthermal
protons, which in turn form a relativistic outflow. SBK demonstrated
that this mechanism can lead to the energization of protons via shear
acceleration in a tenuous corona overlying a hot accretion disk. The
resulting specific energy in the corona can be large enough to drive a
relativistic outflow. In their work, SBK focused on shear acceleration
occurring in a tenuous corona because the density in the underlying
(luminous) disk is so large that collisional losses overwhelm shear
acceleration there. However, in the advection-dominated accretion flows
envisioned here, the low density environment of the (underluminous) disk
itself is expected to be a favorable site for this acceleration
mechanism.

\subsection{\it Model Predictions}

The unique solution for the $\dot M$ variation given by our
equation~(2.21) and the associated density distribution $\rho \propto
(r-2)^{-1}$ have definite consequences for the timing properties of
accretion-powered sources. It has been proposed (Kazanas, Hua \&
Titarchuk 1997; Hua, Kazanas \& Titarchuk 1997; Hua, Kazanas, \& Cui
1999; Kazanas \& Hua 1999) that the timing properties of these sources
(light curves, power spectra, coherence, and time lags between soft and
hard X-rays) can be easily understood as the result of Comptonization of
soft photons produced in the vicinity of the compact object by a hot
($10^8$ K) corona that extends to $r \sim 10^3$ Schwarzschild radii. In
particular, it was pointed out that such an interpretation provides a
means for probing the density structure of the Comptonizing corona,
mainly as a consequence of the dependence of the X-ray hard lags on the
Fourier frequency. Hua, Kazanas, \& Cui (1999) analyzed and fit the lags
from a number of accreting galactic black hole candidate sources. In a
few cases, the lag data were consistent with the density distribution
$\rho \propto r^{-3/2}$ implied by the ADAF model (Narayan \& Yi 1994).
However, in most cases the lag data were consistent with a corona
density profile $\rho(r) \propto r^{-1}$, which is the behavior that is
uniquely determined by our self-similar RADIOS model for $r \gg 2$.
Hence our pseudo-Newtonian, self-similar RADIOS model may provide the
first explanation for a ubiquitous observational conclusion regarding
the apparent radial density variation in accreting compact sources.

\subsection{\it Conclusion}

We have described the overall framework of a pseudo-Newtonian RADIOS
model that associates relativistic outflows with advection-dominated
accretion disks. The self-similar solutions obtained for the physical
properties of the disk extend over several decades in radius and
therefore provide a meaningful description of the global disk structure.
By extending our model to large radii, we are able to constrain the
parameters in the associated Newtonian ADIOS model. Our analysis yields
unique radial dependences for the accretion rate, the density, and the
pressure. For the special case of a torque-free, gasdynamical wind,
which is the focus here, we find that the density varies as $\rho
\propto r^{-1}$ as $r \to \infty$. This density variation is consistent
with that implied by observations of X-ray time lags in accreting
galactic sources. Moreover, the Shakura-Sunyaev $\alpha$-parameter
implied by our model remains nearly constant, independent of the
viscosity model (see eq.~[3.5]). This is consistent with the basic
notion that the shear stress should be roughly proportional to the
pressure. When combined with the fact that our assumed variation
of $\Omega$ is consistent with published numerical simulations, we
conclude that our fundamental scaling assumptions are reasonable.
Nonetheless, these assumptions cannot be fully justified a priori,
and need to be tested by global calculations. 
In our RADIOS model, the strength of the outflow and the value of
the mass loss index $p \equiv d\ln\dot M/d\ln r$ are determined
by the shape of the non-Newtonian potential close to the black hole.
However, it is possible that the behavior $p \to 1/2$ as $r \to \infty$
implied by observations is determined, not by the shape of the potential,
but rather by the details of the mechanism driving the outflow
(e.g., second order Fermi acceleration, magneto-centrifugal forces,
convection).  If the latter is true, then the self-similar
solution presented in this paper, though mathematically attractive,
may have limited physical relevance.  More work on this issue is
needed.

The RADIOS model we have developed can produce highly relativistic
outflows, whereas the outflow is always nonrelativistic in the
corresponding ADIOS model. Furthermore, like ADIOS, our model does not
suffer from the strong sensitivity to the value of the adiabatic index
$\gamma$ that is displayed by the standard ADAF models, which require
$\gamma < 5/3$ (e.g., Narayan \& Yi 1994, 1995a). In contrast to the
results for bipolar ADAF outflows obtained by Igumenshchev \& Abramowicz
(1999), which require $\alpha \sim 1$, Figures~1 and 2 suggest that
in our model outflows can occur for values of $\alpha$ well below
unity. This work may therefore help to establish a physical connection
between advection-dominated accretion flows and the relativistic
outflows possibly emanating from underluminous black hole systems
such as Sgr ${\rm A}^{*}$ (e.g., Falcke 1999; Donea et al. 1999). The
model may also be applicable to the mildly relativistic outflows
observed from galactic black hole candidates (Mirabel \& Rodriguez 1999;
Rodriguez \& Mirabel 1999). However, the nature of the fundamental
acceleration mechanism that energizes the wind particles is still an
open question. One possibility discussed here is second-order Fermi
acceleration driven by the shear of the magnetic field lines. Additional
work needs to be done in this area before the complete disk/wind/jet
scenario can be adequately understood.

The authors thank the anonymous referee for a variety of useful
comments. PAB also gratefully acknowledges the hospitality and
support of NASA/GSFC via a NASA/ASEE Summer Faculty Fellowship.

\eject

\centerline{\bf FIGURE CAPTIONS}
\bigskip

Fig. 1. -- Contour plot of the Shakura-Sunyaev viscosity normalization
parameter $\alpha_0$ (eq.~[3.6]) as a function of $\epsilon$ and
$\Omega_0$. The contour values are $\alpha_0 = 1$ ({\it solid
line}), $\alpha_0 = 0.5$ ({\it dashed line}), and $\alpha_0 = 0.01$
({\it dotted line}). In the region between the solid lines, $\alpha_0$
exceeds unity. Also shown for reference is the shaded region of
self-consistency for our model, given by equations~(3.17). Note that
this region contains values for $\alpha_0$ in
the physically acceptable range $0 < \alpha_0 < 1$.

\bigskip

Fig. 2. -- Sections of the $(\Omega_0 \, , \epsilon)$ parameter space
corresponding to the various constraints given by eqs.~(3.17). The
region above the solid line satisfies $\epsilon > \Omega_0^2/4 \,$, the
region above the dashed line satisfies $\epsilon > (5 \, \Omega_0^2 - 1)
/ 4 \,$, and the region below the dotted line satisfies $\epsilon <
\Omega_0^2 \,$. The curved, shaded area is the intersection of these
constraints and therefore it represents the region of self-consistency
for our model, which is also indicated in Fig.~1.
See the discussion in the text.

\bigskip

Fig. 3. -- Comparison of the pseudo-Newtonian and Newtonian
energy transport rates $\enrate$ (eq.~[2.29]; {\it solid line}) and $\enrate_A$
(eq.~[4.5]; {\it dashed line}) in units of $\epsilon \dot M_0$
as functions of the radius in the disk $r$.
In the inner region, $\enrate$ exceeds $\enrate_A$ due to the steepness
of the pseudo-Newtonian potential.
The two results converge as $r \to \infty$.

\bigskip

Fig. 4. -- Pseudo-Newtonian and Newtonian
accretion rates $\dot M$ (eq.~[2.21]; {\it solid line})
and $\dot M_A$ (eq.~[4.5]; {\it dashed line})
in units of $\dot M_0$
plotted as functions of the radius in the disk $r$.
The accretion rate $\dot M$ has a minimum value at $r=6$,
which is the radius of marginal stability, and the inner
edge of our disk model. At large radii, $\dot M$ approaches the
standard ADIOS result given by $\dot M_A$.

\bigskip

Fig. 5. -- Differential luminosities $dL/dr$ (eq.~[4.11];
{\it solid line}) and $dL_A/dr$ (eq.~[4.12]; {\it dashed line})
emiited into the wind are plotted in
units of $\epsilon \dot M_0$ as functions of the
starting radius in the disk $r$. These two functions correspond
to the pseudo-Newtonian
and Newtonian models, respectively. The power emitted
into the wind is greatest is the pseudo-Newtonian case.

\bigskip

Fig. 6. -- Comparison of the pseudo-Newtonian and Newtonian
differential mass-loss rates $d\dot m/dr$ (eq.~[4.11]; {\it solid
line}) and $d\dot m_A/dr$ (eq.~[4.12]; {\it dashed line})
emitted into the wind,
plotted in units of $\dot M_0$
as functions of the starting radius in the disk $r$. In the
pseudo-Newtonian case, this rate approaches zero close to the
radius of marginal stability, and any remaining material
is swept directly into the black hole.

\bigskip

Fig. 7. -- Pseudo-Newtonian and Newtonian disk Bernoulli
parameters~$B(r)$~(eq.~[4.13]; {\it solid line}) and
$B_A(r)$ (eq.~[4.14]; {\it dashed line})
plotted as functions of the radius in the disk $r$ for
$v_0=0$. In panel
({\it a}), $\epsilon = 0.1$, and both $B$ and $B_A$ remain
positive at all radii. In panel ({\it b}), $\epsilon = 0.3$,
and $B$ changes sign at $r=7.5$, whereas $B_A$ remains negative at
all radii. This difference in behavior is due to the strong
variation of the quasi-Keplerian angular velocity $\Omega$ in the inner region
of the disk in the pseudo-Newtonian model.

\bigskip

Fig. 8. -- Asymptotic Lorentz factors $\Gamma$ (eq.~[4.20];
{\it solid line}) and $\Gamma_A$ (eq.~[4.21]; {\it dashed
line}) for the outflows in
the pseudo-Newtonian and
Newtonian models, respectively, plotted as functions
of the starting radius
in the disk $r$. The Newtonian result is nonrelativistic for
all values of $r$, whereas the pseudo-Newtonian result is
strongly relativistic in the inner region of the disk, and formally
diverges as $r$ approaches the radius of marginal stability
$r_{\rm ms}=6$.

\end{document}